\input harvmac
\lref\sch{J.~H.~Schwarz, {\it Anomaly-Free Supersymmetric Models in Six Dimensions}, hep-th/9512053}
\lref\GS{M.~B.~Green, J.~H.~Schwarz, Phys.~Lett.~{\bf 149B} (1984) 117}
\lref\Sag{A.~Sagnotti, Phys.~Lett.~{\bf 294B} (1992) 196, hep-th/9210127}
\lref\FMS{S.~Ferrara, R.~Minasian, A.~Sagnotti, {\it Low Energy Analysis of M and F theories in Calabi Yau Threefolds}, hep-th 9604097}
\lref\rfs{M.~Bershadsky, V.~Sadov, C.~Vafa, {\it D-Branes and Topological Field Theories,} hep-th/9511222, \hskip 5mm
M.~Douglas, {\it Branes within Branes}, hep-th/9512077, \hskip 5mm
M.~Green, J.~Harvey, G.~Moore, {\it I-Brane Inflow and Anomalous Couplings on D-branes,} hep-th/9605033} 
\lref\mv{D.~Morrison, C.~Vafa, {\it Compactifications of F-Theory on Calabi-Yau Threefolds I,II,} hep-th/9602114, hep-th/9603161}
\lref\sixau{M.~Bershadsky, K.~Intriligator, S.~Kachru, D.~Morrison, V.~Sadov, C.~Vafa,  {\it Geometric Singularities and Enhanced Gauge Symmetries,} hep-th/9606}
\lref\BK{P.~Berglund, S.~Katz, A.~Klemm, P.~Mayr, {\it New Higgs Transitions between Dual N=2 String Models}, hep-th 9605154}
\lref\kmp{S.~Katz, D.~Morrison, R.~Plesser, {\it Enhanced Gauge Symmetry in Type II String Theory,} hep-th/9601108}
\lref\bsv{M.~Bershadsky, V.~Sadov, C.~Vafa, {\it D strings on D manifolds,} Nucl.~Phys.~{\bf B463} (1996) 398, hep-th/9504090}

\Title{\vbox{
\hbox{IASSNS-HEP-96/58}
\hbox{\tt hep-th/9606008}}}
{Generalized Green-Schwarz mechanism in F theory}
\bigskip
\centerline{V. Sadov}
\bigskip
\bigskip\centerline{\it Institute for Advanced Study,
Princeton, NJ 08540}

\vskip .3in
 
We derive the anomaly 8-form of 6-dimensional gauge theories arising in F theory compactifications on elliptic Calabi-Yau threefolds. The result allows to determine the matter content of certain such theories in terms of intersection numbers on the base of elliptic fibration. We also discuss gauge theories on 7-branes with double point singularities on the worldvolume. Applications to Type II compactifications on Hirzebruch surfaces and ${\bf P}^2$ are outlined. 
 
\Date{June 1996}
\vfill\eject  

\newsec{Introduction.}
F theory compactifications on elliptic Calabi-Yau threefolds often contain a gauge theory subsector\mv. To study these gauge theories one can use algebraic methods \sixau,\BK\ which do provide a lot of information. For instance, the gauge group carried by the component of the discriminant is completely determined by the  type of fiber degeneration along that component. It is more difficult to extract the matter content from algebro-geometric data. The complete understanding of how to do this would require better  knowledge of topological theory living on the discriminant. 

The purpose of this paper is to provide {\it topological} rather than algebro-geometric restrictions on the possible matter contents for a given homological type of the discriminant component. These restrictions depend only on the self-intersection number of this component and on its intersection numbers with other components of the discriminant. They are not sensitive to fine geometric details such as singular points. Not surprisingly, they do not fix the matter content completely: for a given homological type there can be many different choices compatible with the restrictions. 
 
In practice it is convenient to combine these restrictions with some local D-brane analysis. Then it is possible to choose certain ``minimal'' matter content for every gauge group which is fixed completely by the topological constraints. For example, for $SU(n)$ theories this ``minimal'' solution includes hypermultiplets in adjoint ${\bf n^2-1}$, fundamental ${\bf n}$ and antisymmetric ${\bf n(n-1)\over 2}$ representations. Such solutions appear to be generic in a sense they correspond to sufficiently generic, smooth 7-brane worldvolumes. 
Some other solutions can be shown to correspond to the situation when the worldvolume develops singularities.

The main technical tool in our analysis is the anomaly 8-form of the gauge theory living on a 7-brane. On the one hand, this 8-form is determined by the matter content of the theory. On the other hand, Green-Schwarz anomaly cancelation mechanism \GS \Sag\ relates it to anomalous couplings on the worldvolume of the 7-brane which can be written explicitly in terms of geometric data. Comparison of these two leads to the topological constraints we are after.       
       
\newsec{Green-Schwarz mechanism in 6 dimensions.}
The anomaly 8-form $I_8$ is defined for any 6-dimensional theory. The actual anomaly
of the action $\delta_\Lambda S=\int I^1_6(\Lambda)$ is related to $I_8$ by the Wess-Zumino descent equation zigzag: 
$$I_8=dI_7,\ \delta_\Lambda I_7=dI^1_6(\Lambda).$$
The 8-form itself is closed and gauge invariant so it can be expressed in terms of characteristic classes of gauge and gravitational field. In notations of Schwarz \sch\
$$\hat{I_8}={10-n\over 8}({\rm tr}R^2)^2+{1\over 6}{\rm tr}R^2\sum_aX^{(2)}_a-{2\over 3}\sum_a X^{(4)}_a+4\sum_{a<b}Y_{ab}$$ 
where
$$X^{(n)}_a={\rm Tr}F^n_a-\sum_R n_{R_a} {\rm tr}_{R_a}F^n_a$$
$$Y_{ab}=\sum_{R_a,R_b'}n_{R_aR'_b}{\rm tr}_{R_a} F^2_a{\rm tr}_{R'_b}F^2_b.$$
For the simple group $G_a$, ${\rm Tr}$ denotes the trace in the adjoint representation, ${\rm tr}_{R_a}$ denotes the trace in the representation $R_a$ and $n_{R_a}$ stands for the number of hypermultiplets in this representation. $n_{R_aR'_b}$ is the number of hypermultiplets in mixed representation $(R_a,R'_b)$ of $G_a\times G_b$. 

Anomaly cancelation via Green-Schwarz mechanism \GS,\Sag~with $n$ antisymmetric tensors requires that $I_8$ be representable as
\eqn\afac{
I_8={1\over 2}\Omega_{ij}X^iX^j,}
where $\Omega_{ij}$ is a bilinear form which is the natural metric on the space of antisymmetric tensors $B^i$, $i=1,\ldots,n$. In turn, the 4-forms $X^i$ are all closed and gauge invariant and can be decomposed  as 
\eqn\fact{
X^i={1\over 2}a^i{\rm tr}R^2 +2b^i_a{\rm tr}F^2_a,}
where ${\rm tr}$ is a trace in a certain preferred representation (this representation is $({\bf n})$ for $SU(n)$). Defined by \afac, $I_8$ can be normalized to set the coefficient of $({\rm tr}R^2)^2$ to ${10-n\over 8}$, as in $\hat{I}_8$. 

Coefficients $a^i,\ b^i_a$ are related to $n_{R_a},\ n_{R_aR'_b}$ by
\eqn\rel{\eqalign{
{\rm index}(Ad_a)-\sum_R{\rm index}(R_a)n_{R_a}&=6(10-n){\Omega_{ij}a^ib_a^j \over \Omega_{ij}a^ia^j} \cr
y_{Ad_a}-\sum_R y_{R_a}\,n_{R_a}&=-3(10-n){\Omega_{ij}b^i_ab^j_a \over \Omega_{ij}a^ia^j} \cr
x_{Ad_a}-\sum_R x_{R_a}\,n_{R_a}&=0\cr
\sum_{R,R'}{\rm index}(R_a)\,{\rm index}(R'_b)\,n_{R_aR'_b}&=(10-n){\Omega_{ij}b^i_ab^j_b \over \Omega_{ij}a^ia^j}, \cr
}}
where we decompose ${\rm tr}_{R_a}F^4=x_{R_a}{\rm tr}F^4 + y_{R_a}({\rm tr}F^2)^2$ assuming $R_a$ has two independent order four invariants. If it has only one, $x_{R_a}=0$.  

Green-Schwarz mechanism \Sag\ of anomaly cancelation amounts to adding to the action $S$ a term
\eqn\cot{A=\int_{{\bf R}^{5,1}} \Omega_{ij}B^iX^j,}
 the gauge invariant field strength 3-form for $B^i$ being $H^i=dB^i+{1\over 2}a^i\omega_{3L}+2b^i_a\omega_{3Y}^a$. As usual, $\omega_{3L}$ is a gravitational Chern-Simons 3-form and $\omega_{3Y}^a$ is a Yang-Mills Chern-Simons 3-form for the $a$-th simple factor. The combination $S+A$ is invariant: 
$$\delta_\Lambda (A+S)=\int \Omega_{ij}(\delta_\Lambda B^i)X^j+\int I^1_6(\Lambda)=0.$$ 

\newsec{Type IIB on a complex surface.}
 
Now let us consider a particular example of 6-dimensional gauge theory coming from F theory compactification on a Calabi-Yau threefold $X$ \mv,\FMS. The appropriate threefold is an elliptic fibration over a complex surface $B$. Over a generic point on the base $B$ the fiber is a smooth elliptic curve.  
Fibers degenerate over a discriminant divisor. The discriminant divisor on $B$  is a union of curves $D_a\subset B$. One can think of compactification of F theory on $X$ as of compactification of Type IIB on $B$. The surface $B$ is not a Calabi-Yau: $c_1(B)\neq 0$. This anomaly is compensated for by adding 7-branes sources located on curves $D_a$. The condition for cancelation of $c_1$ anomaly on $B$ is equivalent \mv~to the Calabi-Yau condition for $X$:
\eqn\canc{
c_1(B)=-{1\over 12}\sum_a N_a\,\delta^{(2)}(D_a)
}
 where the number $N_a$ of 7-branes wrapped around $D_a$ can be determined by the order of vanishing of the discriminant on $D_a$: $N_a={\rm ord}(\Delta)|_{D_a}$. 

The noncompact part of each 7-brane worldvolume is 6-dimensional. If $N_a>1$ there is an enhanced gauge symmetry with gauge group $G_a$ determined by the  geometry of elliptic fibration near $D_a$. This is discussed in \mv~and in great detail in \sixau.

$SU(3)$ holonomy of $X$ implies $h_{20}(X)=h_{10}(X)=0$. In turn, this requires $h_{20}(B)=h_{10}(B)=0$. Thus the only nontrivial Hodge number is $h_{11}(B)=n$.
 Compactifying 10-dimensional self-dual 4-form $B^{(4)}$ on surface $B$ one finds  $n$ antisymmetric 2-tensors $B^i$ in 6 dimensions\FMS:
Let $\{h_i\},\ i=1,\ldots,n$ be a basis of harmonic $2$-forms on $B$ and $\eta_{ij}=\int_B h_ih_j$ be the intersection form. Then a self-dual 4-form is decomposed as $B^{(4)}=B^ih_i$. The scalar product on $B^i$ is thus given by the intersection form $\eta_{ij}$. Among $B^i$ the one coupled to the K\"ahler form on $B$ is self-dual and the other $n-1$ ones coupled to the primitive $(1,1)$-forms are antiselfdual. The scalar product $\Omega_{ij}$ is $(1,\,n-1)$ Lorentz with a signature $\tau(B)=2-n$. 
The 10-dimensional 2-forms $(B^{(2)}_{NS},\,B^{(2)}_{R})$ are frozen at zero values in F theory and do not participate in Green-Schwarz mechanizm. 

It is very natural to expect that any such compactification leads to anomaly-free gauge theory. Thus focusing on this theory itself we expect to find a factorizable 8-form \afac~where the coefficients $a_i,\ b_i^a$ in \fact~are determined by  the geometry of 7-branes. This expectation proves to be true:
Let us decompose 
\eqn\geo{\eqalign{
c_1(B)&=a^ih_i \cr
\delta^{(2)}(D_a)&=b^i_ah_i \cr
}}
The anomaly 8-form  $I_8$ is given by \afac\ with $X^i$  given by \fact~with this particular choice of $a^i$ and $b^i_a$:

\eqn\anf{
I_8={1\over 2}\int_B \big({1\over 2}c_1(B){\rm tr}R^2+2\delta^{(2)}(D_a){\rm tr}(F_a^2)\big)^2
} 

Now let us prove \geo-\anf. The physically meaningful object is the counterterm $A$ \cot. To expose the origin of $A$ it is useful to rewrite it in 10-dimensional terms: 
\eqn\tendim{
 A=\int_{{\bf R}^{5,1}\times B} B^{(4)}\big({1\over 2}c_1(B){\rm tr}R^2+2\delta^{(2)}(D_a){\rm tr}(F_a^2)\big)} 
where we have used \geo\ for $X^i$. 
The effective action \tendim\ is a consequence of a  10 dimensional coupling  $$\int B^{(4)}({1\over 2}{\rm tr}F^2_a -{N_a\over 48}p_1(R))\delta^{(2)}(D_a)$$ 
described in \rfs\
for  $N_a$ coinciding 7-branes wrapped around $D_a$. This coincides with \tendim\ after one uses \canc\ to decompose $c_1(B)$ in terms of $\delta^{(2)}(D_a)$.

In the presence of 7-branes the field strength 5-form for $B^{(4)}$ is $H^{(5)}=dB^{(4)}+{1\over 2}c_1(B)\omega_{3L}+2\delta^{(2)}(D_a)\omega_{3Y}^a$. Note that because of the $c_1$ anomaly cancelation \canc\  
one can rewrite the Bianchi identity as $dH^a=(-{N_a\over 48}{\rm tr}R^2+{1\over 2}{\rm tr}F^2_a)\delta(D_a)$ thus separating individual contributions from different 7-brane worldvolumes. Compactifying $H^{(5)}$ on $B$ one recovers $n$ 3-forms $H^i$ requisite for the Green Schwarz mechanism in section 2. The other two 3-forms $H^{n+1}$ and $H^{n+2}$ are the standard 3-forms from ten dimensions\FMS.

\newsec{Applications and examples.}

\subsec{Matter content of F theory.}

The most important consequence of the above analysis is that we can obtain certain information about matter content of the gauge theory living on  7-branes. In \rel, the left hand side is a particular combination of numbers of hypermultiplets while the right hand side is geometric. It is convenient to represent $c_1(B)$ by the canonical divisor $K$ and $\delta^{(2)}(D_a)$ by divisors $D_a$. Then $\Omega_{ij}a^ia^j=(K\cdot K),\ \Omega_{ij}a^ib_a^j=(K \cdot D_a),\ \Omega_{ij}b_a^ib^j_b=(D_a\cdot D_b)$.  The combination $(K\cdot K)-\tau(B)$ can be written as $8\chi ({\cal O}_B)
=8(1-h_{10}(B)+h_{20}(B))=8$ using the index theorem, so finally $(K\cdot K)=10-n$ and the coefficient in front of $({\rm tr}R^2)^2$ term is  exactly ${10-n\over 8}$ so the anomaly 8-form \anf\ is normalized precisely as $\hat{I}_8$.  
Finally, \rel\ reads 
\eqn\rrel{\eqalign{
{\rm index}(Ad_a)-\sum_R{\rm index}(R_a)n_{R_a}&=6{(K\cdot D_a)
} \cr
y_{Ad_a}-\sum_R y_{R_a}\,n_{R_a}&=-3{(D_a \cdot D_a)} \cr
\sum_{R,R'}{\rm index}(R_a)\,{\rm index}(R'_b)\,n_{R_aR'_b}&={(D_a\cdot D_b)}. \cr
}} 

%
%
Before going into examples, let us make a general remark. Consider the last equation in \rrel, the one that counts mixed $(RR')$ representation. It tells us that every intersection point contributes to the sum by $1$. This looks like an independent check of the local D-brane analysis of \bsv --- every intersection point contributes independently and  the geometry away from the intersection point is irrelevant.   

\subsec{$SU(n)$ and $Sp(k)$ gauge groups on smooth and singular curves.}
Consider a situation
when the discriminant locus has a smooth component $D$ with $A_1$ singularity along it. One expects \mv \sixau\ to find an $SU(2)$ gauge theory on the noncompact part of the worldvolume of 7-brane wrapped around $D$. The number $N_A$ of hypermultiplets in the adjoint ${\bf 3}$ is given by the number $g$ of holomorphic 1-forms on $D$ \kmp,\sixau. This number is given by the adjunction formula $2(g-1)=(D\cdot D)+(K\cdot D)$. We will make a minimal assumption that all other $N_F$ hypermultiplets are in fundamental ${\bf 2}$ and try to solve \rel-\rrel. It turns out this is possible indeed with $N_A=g$ and $N_F=-8(K\cdot D)-2(D\cdot D)$. 

This procedure also works for $SU(3)$ where we find, using $y_{Ad}=9$ and $y_F=1/2$, that again $N_A=g$ and $N_F=-9(K\cdot D)-3(D\cdot D)$. In both cases the condition $N_F>0$ puts constraints on possible choice of $D$.

It was found in \sixau\ that a generic ``nonsplit'' $A_{2k-1}$ singularity on a Hirzebruch surface leads to a gauge group $Sp(k)$ rather than $SU(2k)$. The matter consists of $N_A$ adjoints ${\bf k(2k+1)}$,
$N_F$ fundamentals ${\bf 2k}$ and $N_{AS}$ antisymmetric tensors ${\bf k(2k-1)-1}$. Let us try this matter content for general $B$. Using group-theoretical factors $x_R,\ y_R$ for these representations one does find a consistent solution with $N_A=g$,
$N_{AS}=((D\cdot D)-(K\cdot D))/2$ and $N_F=-8(K\cdot D)-2k(D\cdot D)$.

Gauge groups $SU(n),\ n>3$ appear when there is a ``split'' $A_{n-1}$ singularity along $D$. The matter includes $N_A$ adjoints ${\bf n^2-1}$, $N_F$ fundamentals ${\bf n}$ and $N_{AS}$ antisymmetric tensors ${\bf {n(n-1)\over 2}}$. Solving \rel-\rrel\ one finds: $N_A=g$, $N_{AS}=-(K\cdot D)$ and $N_F=-8(K\cdot D)-n(D\cdot D)$.

One consistency check of the above results is to consider Higgsing $SU(n)\rightarrow SU(n-1)$ using fundamentals and $SU(2k)\rightarrow Sp(k)$ using antisymmetric tensors. A reader can easily convince oneself it works.

An important question arises to what extent the solutions we find is unique. For example, instead of ``minimal'' choice for the matter we made for $SU(3)$ we could try
$N_A$ adjoints ${\bf 8}$, $N_F$ fundamentals ${\bf 3}$ (or $\bar{\bf 3}$) and $N_S$ symmetric tensors ${\bf 6}$. Then the equations \rel-\rrel\ do not fix these numbers completely leaving one free parameter $r$ which can be identified with the number of ${\bf 6}$. The solution is: $N_S=r$, $N_A=g-r$, $N_F=-9(K\cdot D)-3(D\cdot D)+r$. What is the meaning of this solution?

To answer this question one should realize that the relations \rrel\ are essentially {\it topological} and as such sensitive only to homological type of $D$. For instance, they cannot distinguish a singular $D$ from a smooth one simply because {\it generically} all curves are smooth. The solution with $r=0$ corresponds to smooth $D$. It is generic in a sense that other solutions with $r\neq 0$ live in codimension $r$ subspace inside it. For $r>0$ the number of adjoints $N_A=g-r$ does not coincide with the genus computed for the {\it smooth} curve. It leads to a conclusion that the curve $D$ is not smooth in this case, but 
 has $r$ double points instead. (Double points are ``generic singularities'' for curves.) The number of holomorphic 1-differentials counting the adjoints is indeed $g-r$ which shows we are on the right track. Now, every double point 
 locally looks like an intersection of two {\it different} 7-branes with 
 independent $SU(3)$ on each. As such, it should produce a ${\bf (3,3)}$  hyper multiplet.
 But because there is just a single curve $D$ globally, there is only one  diagonal $SU(3)$. With respect to this diagonal $SU(3)$ the representation ${\bf (3,3)}$ breaks as $\bar{\bf 3}+{\bf 6}$. Thus every double point contributes by one hypermultiplet in ${\bf 6}$ and one hypermultiplet in $\bar{\bf 3}$ to the matter content.  But this is exactly what the above formulas for $N_F$ and $N_S$ tell us! 

Generally, an $SU(n)$ gauge theory on $D$ with $r$ double points has $N_A=g-r$ adjoints, $N_{AS}=-(K\cdot D)+r$ antisymmetric tensors, $N_S=r$ symmetric tensors and 
$N_F=-8(K\cdot D)-n(D\cdot D)$ fundamentals.

\subsec{SO(n) gauge groups.}

In F theory the gauge theory with $SO(n), n\geq 7$ appears when there is a D-type singularity along the curve $D$. The index $n$ is even or odd depending on whether this singularity is split or nonsplit on $D$. Generalizing \sixau, we suggest that on smooth generic $D$ the matter consists of $N_A$ adjoints, $N_F$ vectors ${\bf n}$ and $N_S$ spinors. It should be noted that the index of ${\bf n}$ representation is 2, so the symbol ${\rm tr}$ denotes {\it one-half} of the trace in ${\bf n}$. We will use the identities:
\eqn\son{\eqalign{
&{\rm Tr}F^4=2(n-8){\rm tr}F^4+12({\rm tr}F^2)^2,\ {\rm Tr}F^2=2(n-2){\rm tr}F^2,\cr
&{\rm tr}_F F^4=2{\rm tr}F^4,\ {\rm tr}_F F^2=2{\rm tr}F^2,\cr
&{\rm tr}_S F^4=-{d_S\over 8}{\rm tr}F^4+{3d_S\over 16}({\rm tr}F^2)^2,\ {\rm tr}_S F^2={d_S\over 4}{\rm tr}F^2,
}}
where we denoted by $d_s$ the dimension of the appropriate spin representation.

Applying \son\ to \rel-\rrel\ one finds: $N_A=g$, $N_S={16\over d_S}(-2(K\cdot D)-(D\cdot D))$ and $N_F=-{n-4\over 2}(K\cdot D)-{n-6\over 2}(D\cdot D)$. Note that for $SO(8)$, the number of ${\bf 8}_v$ is the same as the number of ${\bf 8}_a+{\bf 8}_c$:  $N_S=N_F=-2(K\cdot D)-(D\cdot D)$ as it should be by triality.

\subsec{Examples: Hirzebruch surfaces and ${\bf P}^2$}

Now let us apply these results to F theory compactification on Hirzebruch surfaces,  discussed in \mv,\sixau. We follow notations of \mv.
For Hirzebruch surface $F_n$ the basis of 2-forms is given by $D_s,\, D_v+{n\over 2}D_s$ with the intersection form given by Pauli matrix $\sigma_1$. The canonical class $K=-2D_s-2(D_v+{n\over 2}D_s)$ has self-intersection 8. The signature $\tau(F_n)=0$ so  the contribution of $(B^{(2)}_{NS}, B^{(2)}_R)$ 
to gravitational anomaly is zero. Thus we expect the anomaly 8-form to factorize: $I_8=({\rm tr}R^2-u\,F^2)({\rm tr}R^2-v\,{\rm tr}F^2)$. 

One finds various gauge groups \sixau\ by wrapping 7-branes around rational curve represented by the divisor $D_u=D_v+nD_s=(D_v+{n\over 2}D_s)+{n\over 2}D_s$. Using \geo\ one obtains $(u,v)=(2,n)$. Applying the above formulas for the matter content of $SU(n)$, $Sp(k)$ and $SO(n)$ gauge theories one recovers results of Table 3 in \sixau. For example, for $SU(2)$ we find $N_{\bf 2}=-8(K\cdot D_u)-2(D_u\cdot D_u)=16+6n$ and $N_{\bf 3}=0$ because $D_u$ has genus zero.  
We did not derive the general formulas for the matter content of theories with exceptional gauge groups, but this can be easily done and the results confer to Table 3 in \sixau\ once applied to the Hirzebruch surfaces.

The only cases when the matter content is not described by the formulas above are
$SU(6)$ with $r/2$ 3-index antisymmetric tensors ${\bf 20}$ and $Sp(3)$ with $r/2$ of ${\bf 14'}$. There is no contradiction, since when one allows for additional $r/2$ ${\bf 20}$ or ${\bf 14'}$ and solves \rel-\rrel\ one recovers the $r$-dependent family of solutions much like it was for $SU(n)$ on a curve with double points. But unlike that case, here the number of adjoints is $r$-independent and equal to $g$. Therefore the origin of these exotic representations is not in singularities of the curve $D$.

One can also consider gauge theories living on curves representable by $p(D_v+{n\over 2}D_s) + qD_s$. Such curves have genus $(p-1)(q-1)$ so the 6-dimensional field theory contains $(p-1)(q-1)$ hypermultiplets in adjoint. The anomaly 8-form factorizes as $I_8=({\rm tr}R^2-2p\,{\rm tr}F^2)({\rm tr}R^2-2q\,{\rm tr}F^2)$. These models were discussed in \sixau\ in the context of F theory compactfication on $F_0={\bf P}^1\times{\bf P}^1$ which is dual to $(12,12)$ heterotic compactification on K3. On the heterotic side $(p,q)$ models provide an example of mixed perturbative non-perturbative gauge symmetry.

Another example is provided by F theory on ${\bf P}^2$. Since $\tau({\bf P}^2)=1$, we expect the anomaly to be
$$I_8={1\over 2}(-{3\over 2}{\rm tr}R^2+2p{\rm tr}F^2)^2.$$

The coefficient $p$ should be interpreted as the degree of the curve $D$ where the gauge group lives. If it is a smooth curve, it has genus $g=p(p-3)/2+1$ which is the number of adjoints in any gauge theory on $D$. For the minimal $SU(2)$ gauge theory the number of fundamentals is $N_F=2p(12-p)$. For example, $p=1$ gives a theory with no adjoints and $22$ fundamentals. It has Higgs codimension $2\cdot 22-3=41$. For $SU(3)$ the number of fundamentals is $N_F=3p(9-p)$. On $p=1$ divisor there are $24$ fundamentals  and the Higgs codimension is $64$. As an independent test, let us obtain these codimensions using algebraic approach of \mv\ and \sixau.  

An elliptic fibration on ${\bf P}^2$ can be described by two homogeneous polynomials
$f_{12}(z_1,z_2)$ and $g_{18}(z_1,z_2)$ which enter the Weierstrass representation
$$y^2=x^3+f_{12}(z)x+g_{18}(z).$$

The $p=1$ curve can be described simply by $z_1=0$. The gauge group $SU(2)$ appears when the discriminant $\Delta(z_1,z_2)=27g_{18}^2-4f^3_{12}$ satisfies $\Delta(0,z_2)=0$ and $\Delta'_{z_1}(0,z_2)=0$. The first equation is solved by $f_{12}(0,z)=3h^2_6(z)$ and $g_{18}(0,z)=2h_6 ^3(z)$. This gives codimension $13+19-7-2=23$. The second equation is solved by $(g_{18})'_{z_1}(0,z)=h_6 (f_{12})'_{z_1}(0,z)$ which produces extra codimension $18$. So the total codimension of $SU(2)$ locus is $23+18=41$ agreeing with the above Higgs codimension. To find $SU(3)$ one needs to set an extra condition
$\Delta''_{z_1z_1}(0,z_2)=0$ which gives additional  codimension 23 adding up with $41$ to $64$.  This verifies our formula for the anomaly $I_8$ for  $\tau(B)\neq 0$.

{\bf Acknowledgments.} The idea of this paper appeared on the final stage of work on \sixau. I am grateful to M.~Bershadsky, S.~Kachru, K.~Intriligator, D.~Morrison, S.~Rey and C.~Vafa for many interesting discussions. This research was supported in part by NSF grants DMS 93-04580 and PHY 92-45317.     

\listrefs

\end